# Challenges in addressing student difficulties with measurement uncertainty of two-state quantum systems using a multiple-choice question sequence in online and in-person classes

Peter Hu, Yangqiuting Li, and Chandralekha Singh
*Department of Physics and Astronomy, University of Pittsburgh, Pittsburgh, PA 15260*

**Abstract**
Research-validated multiple-choice questions comprise an easy-to-implement instructional tool that serves to scaffold student learning and formatively assess students' knowledge. We present findings from the implementation, in consecutive years, of research-validated multiple-choice question sequence on measurement uncertainty as it applies to two-state quantum systems. This study was conducted in an advanced undergraduate quantum mechanics course, in an online and in-person learning environments in consecutive years. Student learning was assessed after receiving traditional lecture-based instruction in relevant concepts, and their performance was compared with that on a similar assessment given after engaging with the multiple-choice question sequence. We analyze and discuss the similar and differing trends observed in the two modes of instruction.

**Introduction**
Measurement uncertainty in quantum mechanics (QM) is a foundational concept that has no classical analogue. Quantum measurement uncertainty is illustrated by the fact that when an observable is measured in a quantum state that is not an eigenstate of the corresponding Hermitian operator, measurement outcomes are not certain. In particular, the measurement collapses the state to one of the eigenstates of the operator corresponding to the observable with a certain probability, and if many measurements of this observable are conducted in an ensemble of identically-prepared systems, the standard deviation of those measurement outcomes is the measurement uncertainty. Furthermore, any subsequent measurements of the same observable made in a collapsed state, assuming no time-evolution, will yield the same outcome with 100% certainty. An observable is said to be well-defined when the quantum system is in an eigenstate of the corresponding operator. Also, the uncertainty principle states that if two observables correspond to operators which do not commute, then they cannot both be measured with 100% certainty, i.e., cannot both be well-defined, in the same quantum state. Because the uncertainty principle can be challenging for students, instructional resources have been developed to help students learn this concept in different situations. Measurement uncertainty and the uncertainty principle are fundamental tenets of quantum theory that are relevant in any context involving successive measurements of different observables, including fields of active research, such as the growing field of quantum information science.

Prior research suggests that students in quantum mechanics courses often struggle with many common difficulties [1–16], including the basic formalism [17,18], notation [19], wavefunctions [20,21], the nature of probability [22], measurement [17,20,23,24], and transferring learning from one context to other contexts [14,25]. For such difficulties as those described, research-validated learning tools can effectively help students develop a robust knowledge structure [26–31]. For example, our group has developed, validated and implemented Quantum Interactive Learning Tutorials (QuILTs) with encouraging results on many topics in QM, including quantum measurement of physical observables [32–34] and the uncertainty principle and Mach-Zehnder interferometer [35,36]. Other commonly used learning tools in physics include clicker questions, first popularized by Mazur [37] using his *Peer Instruction* method. These are conceptual multiple-choice questions presented to a class for students to answer anonymously, typically individually first and again after discussion with peers, and with immediate feedback [37]. They have proven effective and are relatively easy to incorporate into a typical course, without the need to greatly restructure classroom activity or assignments [38].

While these questions can be successfully implemented without additional technological tools, this research used an electronic response system, generally referred to as "clickers," which automatically tracked student responses in real time. When presented in sequences of validated questions, clicker questions can systematically help students with particular concepts that they may be struggling with. Previously, such multiple-choice question sequences, or Clicker Question Sequences (CQS) related to several key QM concepts have been developed, validated and implemented [39–44]. Furthermore, previous work has been conducted to investigate student difficulties with the uncertainty principle as it applies to wavefunctions [15] as well as two-state systems [45], but there has not yet been a documented effort to leverage the CQS method to address those difficulties. Here we describe the development, validation, and implementation of a CQS intended to help students learn about measurement uncertainty as it pertains to two-state quantum systems, and we discuss difficulties in identifying well-defined observables in a given state, calculating measurement uncertainty, successive measurements of various spin angular momentum observables, and other difficulties that naturally came up during implementation.

**Methodology**

*1. Development and validation*

The CQS on quantum measurement uncertainty is intended for use in upper-level undergraduate QM courses. During the development and validation process, we took inspiration from some of the previously-validated learning tools, including determination of learning objectives. In particular, much research involving cognitive task analysis, from both student and expert perspectives, has already been conducted in the development and validation of a QuILT and CQS on measurement uncertainty and the uncertainty principle for wavefunctions (including in the context of the orbital angular momentum), as well as a QuILT on the basics of spin-1/2 systems [35]. Ten student interviews had been conducted using a think-aloud protocol in the development of each of these learning tools, and the insights on student difficulties with regards to spin-1/2 systems and uncertainty principle for orbital angular momentum helped to guide the development of this CQS. In addition, we recently conducted four additional think-aloud interviews with physics post-graduate students with this CQS and, among other things, found that they appreciated the flow of CQS questions in terms of how well they build on each other.

We adapted some of those questions while also drafting and iterating new ones for measurement uncertainty related to two-state quantum systems. To ensure that the material could be completed in the allotted class time, while offering maximal value to students, we prioritized the coverage of conceptual knowledge, used common difficulties as a guide, provided checkpoints that could stimulate useful class discussions, and avoided burdensome calculations. We iterated the questions many times amongst ourselves and with other faculty members to minimize unintended interpretations and ensure consistency and simplicity in terminologies and sentence constructions.

We aimed to address common stumbling blocks and emphasize key features that students may have missed in the large information content of a typical lecture. Some questions in the CQS employ a complex multiple-choice question format, in which students are presented a number of options (e.g., options I, II, III) and must select one of several choices (e.g., choices A, B, C, etc.) that consist of a subset of those options. Though this may increase student cognitive load, we and others in physics education research have successfully used such question models for their relative parsimony in addressing multiple relevant facets of a concept at once, with respect to the limited class time. This is also helped by the immediate feedback and scaffolding support students receive during class discussions of the questions.

The seventeen questions in the final version of the CQS focused on the following four learning goals: identifying well-defined observables in a given quantum state (4 questions), calculations related to measurement uncertainty and the generalized uncertainty principle (5 questions), features and commutation relations of the spin operators $\hat{S}^2, \hat{S}_z, \hat{S}_x$, and $\hat{S}_y$ (5 questions), and compatible vs. incompatible observables (3 questions). (In the CQS, the words "compatible" and "incompatible" are used to describe observables and their corresponding operators interchangeably.)

## 2. Implementation

The data presented here are from administration in a mandatory first-semester junior-/senior-level QM course at a large research university in the United States. Given that a physics course was the object of study, we elected to use quantum physics-framed language rather than quantum information-framed language to discuss these concepts. The final version of the CQS was implemented in two consecutive years, one online and one in person. In each instance, the instructor dedicated between two and three consecutive class sessions, each 50 minutes long, to complete the pre-test, CQS, and post-test.

During the online implementation, the CQS was presented as a Zoom poll while the instructor displayed the questions via the "Share Screen" function. For the in-person implementation, the poll was replaced by a functionally similar classroom clicker system. For each question, the instructor displayed the results after all students had voted, before a full class discussion of the validity of the options provided. Some of the questions involved more calculation than is typical of the conceptual questions in Mazur's method for introductory courses, which are intended to take roughly a minute each. Instructors were urged to use their judgment in giving the class more time to answer such questions.

Because of difficulties in adapting to the online environment in a way that remained conducive to small-group student discussion, the Peer Instruction feature was largely forgone in the online administration, but was realized in full for the in-person administration. We note also that the instructors were different for the online and in-person classes.

Table I below summarizes the learning goals, and the CQS questions and pre-test and post-test questions that cover these concepts.

**Table I.** Details of learning goals and their correspondence to questions on the CQS and the pre-test and post-test.

| Broad learning goals | Details of learning goals | CQS questions covering the concept | Pre-test/post-test questions covering the concept |
|---|---|---|---|
| Identify well-defined observables in a given state | Observables are well-defined in an eigenstate of the corresponding operator | 1.1, 1.2, 1.3, 2.1 | 1, 2, 3, 4a |
| Calculate various quantities related to measurement uncertainty | Definition of uncertainty | 1.4, 2.2 | 1 |
| | Calculating uncertainty | 2.3, 2.4 | 4a, 4b |
| | Generalized uncertainty principle | 1.4, 2.5 | 6 |
| Describe properties of spin operators $\hat{S}^2, \hat{S}_z, \hat{S}_x,$ and $\hat{S}_y$ and their products, as well as features of measurements of the corresponding observables | Commutation relations of spin operators | 3.1, 3.2, 3.3 | 3, 5a-5e |
| | Commuting operators share a complete set of simultaneous eigenstates; non-commuting operators do not | 3.2 3.3, 3.4, 3.5 | 3, 5b, 5c, 5d |
| | Results of measurement of $S^2$ | 3.4, 3.5 | 5e |
| Describe behavior of compatible operators and incompatible operators (in a general sense beyond the spin operators) | Compatible operators share a complete set of simultaneous eigenstates | 4.1 | 4 |
| | Incompatible operators do not share a complete set of simultaneous eigenstates. | 4.2, 4.3 | 4 |

## 3. Assessment

To determine the effectiveness of the CQS, we developed and validated a pre- and post-test containing questions on topics covered in the CQS. The post-test was a slightly modified version of the pre-test, containing changes such as a shift from eigenstates of $\hat{S}_x$ to eigenstates of $\hat{S}_y$, but otherwise remaining conceptually similar. In both online and in-person classes, students completed the pre-test immediately following traditional lecture-based instruction on the topic. After administration of the CQS over two to three class sessions, students completed the post-test. Two researchers graded the pre-test and post-test and, after discussion, converged on a rubric, for which the inter-rater reliability was greater than 95%.

During both online and in-person classes, the two to three classes were dedicated exclusively and consecutively to administration of the pre-test, CQS and ensuing discussions, and post-test. Since the CQS questions build on each other, they typically take less time to complete than if the questions were not part of a sequence. The only other content assigned over this duration was a traditional textbook homework set which overlapped with this period; given our past experience, we do not believe this had a significant effect on the more conceptual post-test performance. That said, this study is nonetheless quasi-experimental [46] in design, in light of these and other factors over which we did not have complete control.

The pre- and post-test questions are reproduced in Appendix A. Questions Q1-Q3 on the pre- and post-test provided students three possible answers from which to choose, and credit was awarded for correctly selecting or omitting each answer, for a total of up to three points per question. For these three questions, correct answers are bolded. For the free-response questions, Q4a was out of one point, while Q4b and all parts of Q5 were scored with two points, one each for answer and reasoning. A more detailed breakdown of the questions is provided in the next section.

**Results**

The pre-test and post-test results for each question, as well as normalized gains [47] and effect sizes [48], corrected for small sample size through a multiplicative factor equal to $\frac{N-3}{N-2.25}\sqrt{\frac{N-2}{N}}$, are listed in Tables II (online with $N = 27$) and III (in-person with $N = 23$). Overall, the results are encouraging, and students performed well on the post-test, with relatively high normalized gains and generally medium to large effect sizes. The multiple-choice questions (Q1-Q3) had reasonably high pre-test scores across both classes.

**Table II.** Results of the online administration of the CQS via Zoom. Comparison of pre- and post-test scores, along with normalized gain [47] and effect size as measured by Cohen's $d$ [48], for students who engaged with the CQS ($N = 27$).

| Question # | Pre-test mean | Post-test mean | Normalized gain | Cohen's $d$ |
|---|---|---|---|---|
| **Q1** | 78% | 94% | 0.72 | 0.86 |
| **Q2** | 94% | 98% | 0.60 | 0.31 |
| **Q3** | 95% | 93% | - | - |
| **Q4a** | 78% | 85% | 0.33 | 0.18 |
| **Q4b** | 52% | 76% | 0.5 | 0.67 |
| **Q5a** | 87% | 93% | 0.43 | 0.20 |
| **Q5b** | 69% | 81% | 0.41 | 0.37 |
| **Q5c** | 70% | 83% | 0.44 | 0.35 |
| **Q5d** | 37% | 76% | 0.62 | 0.85 |
| **Q5e** | 31% | 48% | 0.24 | 0.38 |

**Table III.** Results of the in-person administration of the CQS. Comparison of pre- and post-test scores, along with normalized gain and effect size as measured by Cohen's $d$, for students who engaged with the CQS ($N = 23$).

| Q# | Pre-test mean | Post-test mean | Normalized gain | Cohen's $d$ |
|---|---|---|---|---|
| **Q1** | 80% | 99% | 0.93 | 0.98 |
| Q2 | 91% | 97% | 0.69 | 0.29 |
| Q3 | 83% | 94% | 0.67 | 0.36 |
| Q4a | 65% | 96% | 0.89 | 0.81 |
| Q4b | 35% | 78% | 0.66 | 1.01 |
| Q5a | 65% | 96% | 0.88 | 0.91 |
| Q5b | 52% | 85% | 0.68 | 0.78 |
| Q5c | 46% | 80% | 0.64 | 0.76 |
| Q5d | 48% | 70% | 0.42 | 0.47 |
| Q5e | 50% | 80% | 0.61 | 0.74 |

Table IV presents a brief overview of the CQS data.

**Table IV.** Summary of notable result trends in CQS responses across both years. For ease of comprehension, correctness rates for questions are referred to as low (0-40%), medium (41-70%), and high (71-100%). Note that a small number of questions administered in the in-person implementation did not appear in the online implementation, and vice versa.

| CQS # | Overall performance | Comments |
|---|---|---|
| 1.1-1.4 | Low to medium | Students were only moderately successful at identifying well-defined observables, especially for less straightforward cases, e.g., in CQS 1.2.<br><br>For CQS 1.3, the most popular distractor was the one that asserted that energy could not be well-defined for any state (where the Hamiltonian $\hat{H} = C(\hat{S}_x + \hat{S}_z)$).<br><br>For CQS 1.4, many students considered $\frac{\hbar}{2}$ to be the minimum product of uncertainties for any observables whose operators do not commute; some students also stated that quantum measurement uncertainty comes from apparatus imprecision. |
| 2.1-2.5 | Generally low to medium<br><br>CQS 2.1 had medium performance online and high performance in-person | CQS 2.1 asked students whether uncertainties of specific observables were zero or nonzero, which most answered with reasonable success.<br><br>For CQS 2.2, it was clear that most students knew the formulation $\sigma_A^2 = \langle A^2 \rangle - \langle A \rangle^2$, but were less comfortable with $\sigma_A^2 = \langle (A - \langle A \rangle)^2 \rangle$.<br><br>CQS 2.5 showed that students may have been unfamiliar with applying the generalized uncertainty principle to the spin operators, e.g., $\hat{S}_x$ and $\hat{S}_y$. |
| 3.1-3.5 | Generally high<br><br>Low for CQS 3.3 | Students had an easy time with commutation relations of simple operators (e.g., $\hat{S}_x$), but tended to struggle somewhat more as the operators became more complicated (e.g., $\hat{S}_x \hat{S}_y$). |

| | | For CQS 3.3 online, the instructor gave a hint reminding students of how to simplify commutation relations when operators are multiplied together. During the second vote, the correctness rate improved to medium. |
| | | CQS 3.4-3.5 (in-person only) were meant to help students with the results of a measurement of $S^2$. |
| 4.1-4.3 | Medium to high | CQS 4.1, dealing with properties of observables corresponding to compatible operators, was more challenging for the students than the other questions in this cluster. |
| | | 4.2 (incompatible operators) and 4.3 (generalization of incompatible operators) had high correctness rates; these questions may have been more intuitive or less complicated. |

Below, we discuss some difficulties that were successfully addressed during the administration of the CQS for both years, as well as some that remained for smaller percentages of students.

## *1. Difficulties that were successfully addressed*

### *1.1 Identifying observables that are well-defined in a state*

Questions Q1-Q3 on the pre- and post-test asked students to identify observables that are well-defined in a given state. Most students correctly selected $S_x$ or $S_y$, identifying that the state is an eigenstate of the corresponding operator. In Q1, some students did not select $S^2$, which is also well-defined in the given state because its corresponding operator is proportional to the identity operator and commutes with $\hat{S}_x$. Some students incorrectly selected $S_z$, which may be due to the use of $S_z$ in class as a frequent example where the Hamiltonian $\hat{H} \propto \hat{S}_z$. Questions such as CQS 1.1 and CQS 2.1 address these issues. On the post-test, the correctness rate on questions Q1-Q3 increased, with the exception of Q3 in the in-person administration, which had an especially high pre-test score (see Tables II-III). Since pre-test scores on all three questions were quite high across both years, the normalized gains and effect sizes are not as informative, but this indicates that students have a relatively strong grasp of the concepts involved in these questions. This being said, Q1 had a large effect size for both implementations, indicating noticeable improvement.

### *1.2 Calculating the measurement uncertainty*

On the pre- and post-test, Q4 presented students with two states (Q4a and Q4b). The question asked them to determine whether the uncertainty in measuring a particular component of spin was zero in the given state (similar states involving $x$- or $y$-components of spin were provided in the pre- and post-test, respectively), and to calculate the uncertainty if it was not zero in the given state. Q4a provided a state in which the observable could be measured with 100% certainty, so students who indicated this received full credit. Across both implementations for Q4b, we decided to give full credit to students who were able to provide the formulas for calculating uncertainty (e.g., an observable $A$ has uncertainty $\sigma_A = \sqrt{\langle A^2 \rangle - \langle A \rangle^2}$ and the symbols under square root are the expectation values of $A^2$ or $A$). As the approach for solving the problem, and not the answer of the numerical calculation, was the primary learning goal assessed by this question, our rubric thus avoided penalizing students who made a mathematical error in subsequent steps. Also, some students correctly identified whether the measurement uncertainty is zero or non-zero, but justified their answer only by invoking the probabilities of measuring each possible outcome. These students received half credit based upon our rubric. Questions CQS 2.3 and 2.4 were added for the in-person implementation to address issues related to the calculation of measurement uncertainty, and while Q4a had a high effect size in only the in-person implementation in part due to lower pre-test scores, there were reasonably impressive normalized gains and effect sizes for Q4b across both years (see Tables II-III).

*1.3 Results of successive measurements of identical or non-commuting observables*

On the post-test, Q5 asked students for the final outcome of consecutive measurements of some permutation of the observables $S_x$, $S_y$, and $S^2$, specifically testing whether students could recognize what happens when the measurements involved in the question corresponded to operators that did or did not commute with each other. Question Q5a asked about two consecutive measurements of $S_x$, Q5b about consecutive measurements of $S_x$ and $S_y$, and Q5c about consecutive measurements of $S_x, S_y$ and $S_x$ again. For Q5a, while most students correctly answered that the measurement in the collapsed state will yield $-\frac{\hbar}{2}$ as the outcome, some students did not recognize that $\hat{S}_x$ and $\hat{S}_y$ do not commute. Thus, for question Q5b, they answered that the outcome of the $S_y$ measurement would still be $-\frac{\hbar}{2}$, corresponding to $|-y\rangle$, and that in Q5c, the final $S_x$ measurement would yield $-\frac{\hbar}{2}$, corresponding to $|-x\rangle$, neither of which is correct. Questions such as CQS 2.5 addressed these issues, and in general, the post-test scores show reasonable improvement for all parts of Q5 during the online and in-person administrations (see Tables II-III).

*1.4 Results of successive measurements of $S_x$ and $S^2$*

Question Q5d asked students for the final outcome of consecutive measurements of $S_x, S^2$ and $S_x$ again. On the pre-test, some students stated that for the final measurement of $S_x$, either eigenvalue $\pm\frac{\hbar}{2}$ and eigenstate $|\pm x\rangle$ could be obtained, and some explicitly cited $\hat{S}_x$ and $\hat{S}^2$ not commuting with each other for their reasoning. CQS 3.5 addresses measurement of $S^2$ immediately after $S_x$, and CQS 4.3 and other questions in that sequence helped students generalize from spin-1/2 systems to more generic observables that correspond to operators that do or do not commute, and how such relationships may affect the measurements of those observables in a given quantum state. Student post-test performance showed better understanding of these concepts.

*1.5 Conflation of eigenstates and eigenvalues*

In the in-person implementation, on some questions only on the pre-test (not the post-test), some students wrote, e.g., that "The state will collapse into $-\frac{\hbar}{2}$," or "It is in the state $-\frac{\hbar}{2}$." While the CQS did not explicitly focus on distinguishing between the collapsed state and measured value of the observable, it is likely that the precise language used throughout the CQS helped students distinguish between eigenvalues and eigenstates on the post-test.

**2. Examples of persistent difficulties**

*2.1 Incorrect answers for the results of a measurement*

Across both years' implementations, there were some students who on both the pre- and post-test answered that the result of a measurement was the eigenvalue multiplied by the eigenstate. For example, for question Q5a-e, they stated that making a measurement of $S_x$ in the state $|-x\rangle$ would yield an outcome of $-\frac{\hbar}{2}|-x\rangle$. They also claimed, e.g., that the outcome for Q5a would be $\frac{\hbar^2}{4}|-x\rangle$, the outcome for Q5b would be $\pm\frac{\hbar^2}{4}|\pm y\rangle$, and the outcome for Q5c outcome would be $\pm\frac{\hbar^3}{8}|\pm x\rangle$. This type of reasoning may be closely related to the student difficulty that an operator's action on a quantum state represents a measurement of the corresponding observable in the state and should be investigated further [15]. It is interesting that these students' answers remained the same on the pre-test and the post-test. This difficulty did not fall within the scope of this CQS, but it was addressed in another CQS which will be reported upon in a future publication. This may explain why the difficulty was observed to be rare but resistant for the duration of this CQS.

In the in-person implementation, another mistake was observed somewhat frequently even on the post-test. Question Q5d asked for the final outcome when $S_x, S^2$ and then $S_x$ were measured in that order in immediate succession, with the first measurement of $S_x$ yielding $-\frac{\hbar}{2}$. Some students correctly stated that the intervening measurement of $S^2$ did not collapse the state since the state was already an eigenstate of $\hat{S}^2$, but chose the *wrong* eigenvalue and eigenstate ($+\frac{\hbar}{2}$, corresponding to $|+x\rangle$). These students were given full credit in recognition of their correct reasoning. If this is reflective of a deeper difficulty rather than a careless mistake, we have no compelling speculation with regard to what that difficulty may be.

*2.2 Conceptual difficulties and ambiguities regarding uncertainty relations*

In the in-person implementation, we added an additional free-response question (not shown in Table III) asking students to explain in their own words what it means for two observables, A and B, to have an uncertainty relation between them. Most students gave an answer involving the inability to know with full certainty the values of observables whose corresponding operators do not commute, but some framed their answers in terms of the position-momentum uncertainty relation (commonly cited as $\sigma_x \sigma_{p_x} \geq \frac{\hbar}{2}$), which is not related to two-state systems, rather than using the generalized uncertainty principle. As a result, these students noted that the measurement uncertainties of two observables must multiply to be greater than or equal to $\frac{\hbar}{2}$. These results support the findings of a previous study [45]. Though many questions in the CQS implicitly required knowledge of an uncertainty relation, only CQS 2.5 explicitly discussed the uncertainty principle and its applications to spin-1/2 systems, so this concept is worth emphasizing more in the future.

Another related response to this broad question about the uncertainty principle was "You can measure A or B, but not both." This type of response was observed only in the pre-test, and not the post-test, indicating that these students may have realized the difference between the ability to physically measure an observable versus being able to predict with 100% certainty what outcome would be obtained when the measurement of an observable is made. Other students were not strictly incorrect, but were somewhat unclear in their qualitative responses, which included statements such as "If we know the value of A, then B is a point of complete uncertainty," or "if we know A with 100% certainty, we will have 0% certainty for B." One interpretation of these responses is that "complete uncertainty" or "0% certainty" would refer to, in the case of a two-state system, an equal 50% chance of measuring either outcome. However, it is also possible that these students had the position-momentum uncertainty relation in mind, in which it is intuitive that a (continuous) decrease in uncertainty $\sigma_x$ must be compensated for by an increase in uncertainty $\sigma_{p_x}$ in order to maintain the product to be greater than or equal to $\frac{\hbar}{2}$. In the case of a two-state system, rather than refer to this as "complete" uncertainty, a more accurate description may be something like "maximum" uncertainty. (It is worth noting that, in a spin-1/2 system, the maximum uncertainty for any component of spin is exactly $\frac{\hbar}{2}$.)

Furthermore, in response to this question about the uncertainty principle between observables A and B, many students stated something more general to the effect of "we can never know both quantities exactly at the same time" if their corresponding operators did not commute. Additionally, many students noted that "measurement of one [observable] will possibly affect the other," referencing the collapse to an eigenstate of the observable measured, in which the other observable would not be well-defined. Students were given full credit for all of these responses, since they have all articulated that the more is known about the value of one observable, the less is known about the value of another observable when their corresponding operators do not commute. While on the pre-test, some students were confused about uncertainty relations or left the question blank, nearly every student answered on the post-test in one of the ways discussed and thus earned full credit.

### *3. Comparisons between online and in-person implementations*

Before comparing the online and in-person implementations, we note that some revisions were made to improve the CQS. These improvements affected the presentation of concepts that are covered in questions Q3, Q4b, and Q5d-e, mostly by providing more scaffolding to help students with solving these problems. The pre- and post-test questions remained nearly unchanged, with the exception of a clarification for questions Q5a-e as described in the next section. A one-to-one comparison can thus be drawn between the online and in-person implementations for pre- and post-test questions Q1, Q2, Q4a, and Q5a-c, as the CQS questions that covered the relevant concepts did not undergo any changes between years. For these questions, students' post-test performance does not differ appreciably between years. However, for the free-response questions Q4a and Q5a-c, the gap between pre-test and post-test performance is larger for the in-person class, as indicated by the larger normalized gains and effect sizes (see Tables II-III).

For all questions on the pre- and post-test, the trend appears to remain similar: The average pre-test performance in the in-person implementation was, in general, somewhat lower, but post-test performance for both groups was comparable for all questions aside from Q5e, which is discussed in the next section. It is interesting that students performed about equally well on the post-test for both administrations, given that the online learning environment had greatly reduced opportunity for peers to discuss their responses with each other. We acknowledge that one possibility is students' ability to consult resources, despite being instructed not to do so, during the online-administered pre- and post-tests. Even though students were told that the quizzes were closed-book and closed-notes, such a rule could not be enforced when, as was the case, most students had their cameras off. Even so, those students would have had access to the same resources during both the pre- and post-test, so the sizable improvements in the post-test scores of the online class are still a good sign of the benefits of the CQS. We also observed that many more students left some pre-test questions completely blank in the in-person implementation compared to the online implementation, despite both classes having received the same amount of time to complete the pre- and post-tests. Since the students were given sufficient time to complete the pre- and post-tests, the cause of this higher occurrence of leaving some questions unanswered during the in-person class is unclear. This may be due to students not feeling confident enough to answer, or dealing with additional test anxiety or apathy not experienced by the students in the online implementation. In particular, while online classes have their disadvantages, there were also some benefits and conveniences that would have been lost in the transition back to in-person classes, which could have contributed to this phenomenon. Finally, as we have noted before, the different instructors between years could have also been a factor; e.g., during the online administration, it is possible that more emphasis was placed on content related to two-state spin systems.

The administration of the CQS in an online learning context may have affected student performance differently as compared to the in-person administration. However, regardless of whether the performance across years can be compared one-to-one, it is clear that the CQS has had a beneficial impact on student learning for both the online and in-person implementations.

*3.1 Result of measurement of $S^2$*

For the in-person implementation, CQS 3.4-3.5 were added to provide additional scaffolding on concepts relating to the observable $S^2$. Although the two implementations therefore cannot directly be compared with regard to student learning of these concepts, there are some differences worth examining. During the online implementation of the CQS, students demonstrated difficulties with the outcomes obtained from a measurement of $S^2$. This is relevant in post-test question Q5e, which asked students to provide the outcomes of successive measurements of $S_x, S_y$, and $S^2$ in a state, in that order. During online administration, this question had a large variety of responses that were difficult to score, but were still useful in shedding light on student difficulties. In particular, some students did not recognize that the eigenvalues obtainable from a measurement depend on the observable being measured. As an example, $\hat{S}_z$ and $\hat{S}^2$ share eigenstates, but a measurement of $S_x$ made in the state $|-x\rangle$ would yield $-\frac{\hbar}{2}$, whereas a measurement of $S^2$ in that same state would yield $\frac{3}{4}\hbar^2$. However, multiple students answered that, should the measurements of $S_x$ and $S^2$ be made in succession in this state, they would both yield eigenvalues of $-\frac{\hbar}{2}$. One possible reason for this type of response is that these students had associated the label $|-x\rangle$ primarily with the eigenstate of operator $\hat{S}_x$ with eigenvalue $-\frac{\hbar}{2}$, so although some students realized that this state is a simultaneous eigenstate of $\hat{S}^2$, they had difficulty with the corresponding eigenvalue (the state $|-x\rangle$ also carries an implicit label for the quantum number $s$, which is dropped for ease of notation, which may contribute to this difficulty).

Question Q5e also revealed that some students may not have realized when answering the question that the only possible measured eigenvalue of $S^2$ in any two-state spin system is $\frac{3}{4}\hbar^2$, as the only eigenvalue of $\hat{S}^2$ is $\hbar^2 s(s+1)$, with $s = \frac{1}{2}$ for spin-1/2 systems. When listing the possible eigenvalues for the measurement of $S^2$, in addition to the previously-mentioned $-\frac{\hbar}{2}$ (and the closely related $\frac{\hbar}{2}$), other answers included $\pm\frac{\hbar^2}{4}$ and $-\frac{3}{4}\hbar^2$, which are not unreasonable responses. The response $\pm\frac{\hbar^2}{4}$ may come from literally squaring the eigenvalues of, e.g., $\hat{S}_x$ while matching the sign of the eigenstate label $|\pm x\rangle$, in analogy with the eigenvalues of $\hat{S}_x \pm\frac{\hbar}{2}$ being associated with the respective eigenstates $|\pm x\rangle$. Finally, the response $-\frac{3}{4}\hbar^2$ may be analogous to the notion that the state $|-x\rangle$ is associated with a negative eigenvalue $-\frac{\hbar}{2}$ for a measurement of $S_x$.

In the online administration, the CQS did not explicitly address measurements of the observable $S^2$. Therefore, it is not surprising that students were left with some alternative conceptions. As a result of this, we refined the CQS to explicitly address the eigenvalues and eigenstates obtained from a measurement of $S^2$ in the following in-person implementation. On Q5a-e of the pre- and post-test for the in-person implementation, we made the small addition of explicitly asking students to provide both the eigenvalue and eigenstate resulting from the measurements posed, instead of simply asking "What is the result of the measurement?" as in the online administration in the preceding year (see Appendix A for details). Previously, for the online implementation, students were given credit for providing either the obtained eigenvalue or the eigenstate after the measurement. During the in-person implementation, as a result of the change, students provided less ambiguous answers to these questions. Though the differences in post-test performances for online and in-person administrations could be attributable to more than one factor, the additional scaffolding provided by CQS 3.4-3.5, which specifically addressed these concepts, appears to have been effective. Across the two years, students' performance rose from 48% percent in the online implementation to 80% on the in-person implementation (see Tables II-III).

A summary of student difficulties observed in the pre- and post-tests is presented in Table V.

**Table V.** Student difficulties addressed by the CQS questions, which are found in Appendix B.

| Concept | Observed student difficulties | CQS # | Pre-/post-test # \| comments |
|---|---|---|---|
| Identifying observables that are well-defined in a state | Not realizing that an eigenstate of e.g., $\hat{S}_y$ is also an eigenstate of $\hat{S}^2$; stating that $\hat{S}_z$ is well-defined in an eigenstate of $\hat{S}_y$ | 1.1, 1.2, 1.3, 2.1, 2.5 | Q1, Q2, Q3, Q4a \| Some improvement |
| Calculating measurement uncertainty | Using nonzero probability of measuring each outcome to justify nonzero uncertainty rather than actually calculating uncertainty | 2.2, 2.3, 2.4, 2.5 | Q4b \| Major improvement |
| Results of successive measurements of observables whose corresponding operators are compatible or incompatible | *[For in-person, many students did not provide both the eigenvalue and the state on the pre-test, but did provide both on the post-test]* | 3.1, 3.2, 3.3, 4.1 | Q5a \| Some improvement for online; major improvement for in-person |
| | Stating that successive measurements of incompatible observables (e.g., $S_x$ and $S_y$) will yield the same eigenvalue or not change the state | 3.1, 3.2, 3.3, 4.2, 4.3 | Q5b, Q5c \| Some improvement for online; major improvement for in-person |
| | Stating that successive measurements of compatible observables (e.g., $S^2$ and $S_x$) will change the state | 3.1, 3.2, 3.3, 3.4, 3.5, 4.1 | Q5d \| Major improvement for online; some improvement for in-person |

| Conflation of eigenstates and eigenvalues | Stating, e.g., that a measurement will "collapse the state into $-\frac{\hbar}{2}$" | 3.4, 3.5 | Q5a-e \| Major improvement (in-person only) |
|---|---|---|---|
| Conceptual difficulties regarding uncertainty relations | Stating, for example, that if operators $\hat{A}$ and $\hat{B}$ (corresponding to observables $A$ and $B$) do not commute:<br>• $\sigma_A \sigma_B \geq \frac{\hbar}{2}$ for generic observables $A$ and $B$<br>• Observables $A$ and $B$ cannot both be measured (often unclear whether referring to the measurements themselves or the precision of the measurements)<br>• Knowing $A$ implies infinite uncertainty for $B$ | 1.4, 4.3 | Q6 \| Major improvement (in-person only) |
| Result of measurement of $S^2$ | Giving an incorrect eigenvalue for a measurement of $S^2$ | 3.4, 3.5 | Q5e \| Some improvement for online; major improvement in-person |

**Summary**

Validated CQS can be effective tools when implemented alongside classroom lectures. We developed, validated, and found encouraging results from implementation of a CQS on the topic of measurement uncertainty in two-state quantum systems, in both online and in-person settings. Post-test scores improved for every question following the administration of the CQS, with the exception of Q3 in the online implementation; this question tested students on whether energy can be well-defined for various Hamiltonians, and had an exceptionally high pre-test score. While the performance on the multiple-choice questions was high to begin with on the pre-test, there was significant improvement in the free-response questions. Effect sizes varied for the online implementation, but notably were large for nearly every free-response question in the in-person implementation, since the pre-test scores were lower than in the online implementation. This difference in pre-test performance could be reflective of differences in student preparation or behavior, or other factors such as instructor or environment. An examination of the post-test scores in the online and in-person implementations shows comparable performance in both years on most questions, demonstrating the effectiveness of the CQS in both administrations.

**Ethical statement**

**Acknowledgments**
We thank the NSF for award PHY-1806691. We thank all students whose data were analyzed and Dr. Robert P. Devaty for his constructive feedback on the manuscript.

**Appendix A: Pre-test and post-test questions**

*The states associated with the z-component of spin with eigenvalues $\frac{\hbar}{2}$ and $-\frac{\hbar}{2}$ are $|+z\rangle$ and $|-z\rangle$, respectively (similar notation is used for the x- and y-components). All other notations are standard.*

*Additionally, students were given the following notes:*
- *The spin operators $\hat{S}_z$, $\hat{S}_x$, and $\hat{S}_y$ correspond to the observables $S_z$, $S_x$, and $S_y$, respectively, which in turn correspond to the z-, x-, and y-components of a spin-1/2 particle's spin. All notations are conventional.*
  - $\hat{S}_z|z\rangle = \frac{\hbar}{2}|+z\rangle$      $\hat{S}_z|-z\rangle = -\frac{\hbar}{2}|-z\rangle;$
  - $\hat{S}_x|x\rangle = \frac{\hbar}{2}|+x\rangle$      $\hat{S}_x|-x\rangle = -\frac{\hbar}{2}|-x\rangle$
  - $\hat{S}_y|y\rangle = \frac{\hbar}{2}|+y\rangle$      $\hat{S}_y|-y\rangle = -\frac{\hbar}{2}|-y\rangle$
  - $\hat{S}^2|\pm\alpha\rangle = \frac{3\hbar^2}{4}|\pm\alpha\rangle$ where $\alpha = x, y, z$
- *Measurement in a state in which an observable is well-defined will yield a particular eigenvalue with 100% certainty (i.e., the state is an eigenstate of the operator corresponding to the observable).*

*Correct answers for multiple-choice questions are bolded.*

1. Consider a system in the state $|+x\rangle$. Which of the following observables are well-defined (i.e., they can be measured with no uncertainty) in this state?
   I. **$S^2$**
   II. **$S_x$**
   III. $S_z$

2. Consider a system in the state $|+x\rangle$. Measurements of which of the following observables would yield a well-defined value?
   I. Energy, if the Hamiltonian is $\hat{H} = C\hat{S}_z$ (where $C$ is an appropriate constant)
   II. **Energy, if the Hamiltonian is $\hat{H} = C\hat{S}_x$ (where $C$ is an appropriate constant)**
   III. **Any observable whose corresponding operator commutes with $\hat{S}_x$**

3. Consider the Hermitian operators $\hat{A}$ and $\hat{B}$, which correspond to observables. They are *incompatible* operators. The Hamiltonian is given by $\hat{H} = \hat{A} + \hat{B}$. Suppose you measure energy and obtain $E_0$. Choose all of the following statements that are correct immediately after the measurement of energy.
   I. A is well-defined.
   II. B is well-defined.
   III. The state collapses to an eigenstate of $\hat{B}$ immediately after the measurement of energy.

4. For the following states of a system, does a measurement of the observable $S_y$ yield a value with 100% probability? If the uncertainty is non-zero, calculate it.

a. $\frac{1}{\sqrt{3}}|+y\rangle + \sqrt{\frac{2}{3}}|-y\rangle$
b. $|+y\rangle$

5. Consider the Hermitian operators $\hat{S}_x$, $\hat{S}_y$, and $\hat{S}^2$. Suppose you made a measurement of the observable $S_x$ for a system in some state, and obtained the value $-\frac{\hbar}{2}$.

   a. Suppose you immediately made another measurement of the observable $S_x$. What are the possible values that you can measure? What is the state immediately after the measurement? Explain.

   b. Suppose you instead immediately measured the observable $S_y$ after the first measurement of $S_x$. What are the possible values that you can measure? What is the state immediately after the measurement? Explain.

   c. Suppose after the first measurement of $S_x$, you measured $S_y$ in immediate succession, and $S_x$ once again. What are the possible values that you can measure? What is the state immediately after the measurement? Explain.

   d. Suppose after the first measurement of $S_x$, you measured $S^2$ in immediate succession, and $S_x$ once again. What are the possible values that you can measure? What is the state immediately after the final measurement? Explain.

   e. Suppose after the first measurement of $S_x$, you measured $S_y$ in immediate succession, and then $S^2$ immediately after that. What are the possible values that you can measure? What is the state immediately after the final measurement? Explain.

6. [In-person only] What does it mean for two observables in QM to have an uncertainty relation? Explain in your own words, and imagine your audience to be a high school student.

## Appendix B: Clicker questions
*When the state $|\chi\rangle$ appears in the clicker questions, it refers to a generic state.*

**CQS 1.1** Consider a system with a Hamiltonian $\hat{H} = C\hat{S}_x$, where $C$ is an appropriate constant. Choose all of the following statements that are correct for a system in an eigenstate of $\hat{S}_z$, i.e., $|\pm z\rangle$.
I. The observable $S_x$ is well-defined.
II. The observable $S_z$ is well-defined.
III. Energy is well-defined.
A. I only     **B. II only**     C. II and III only     D. I and III only     E. None of the above

**CQS 1.2** Consider a system with a Hamiltonian $\hat{H} = C\hat{S}_x$, where $C$ is an appropriate constant. Choose all of the following statements that are correct for the state $\frac{1}{\sqrt{2}}(|+z\rangle + |-z\rangle)$.

I. The observable $S_x$ is well-defined.
II. The observable $S_z$ is well-defined.
III. Energy is well-defined.

A. I only   B. II only   C. III only   **D. I and III only**   E. None of the above

**CQS 1.3** Consider a system with a Hamiltonian $\hat{H} = C(\hat{S}_x + \hat{S}_z)$, where $C$ is an appropriate constant. Choose all of the following statements that are correct for the state $\frac{1}{\sqrt{2}}(|+z\rangle + |-z\rangle)$.

I. Since $[\hat{S}_x, \hat{S}_z] \neq 0$, energy cannot be well-defined in *any* state of this system.
II. The observable $S_x$ is well-defined.
III. The observable $S_z$ is well-defined.

A. I only   **B. II only**   C. III only   D. I and II only   E. None of the above

*Class discussion for CQS 1.1-1.3*
- When an observable is well-defined in a state, measurements of this observable will return a predictable result with 100% certainty.
- The particular state $\frac{1}{\sqrt{2}}|+z\rangle + \frac{1}{\sqrt{2}}|-z\rangle$ is an eigenstate of $\hat{S}_x$, but $S_x$ would not be well-defined for arbitrary coefficients of $a|+z\rangle + b|-z\rangle$.
- Discuss that the Hamiltonian $\hat{H} = C(\hat{S}_x + \hat{S}_z)$ is perfectly acceptable: not only is it diagonalizable and has its own eigenstates (which are not the same as those of either $\hat{S}_x$ or $\hat{S}_z$), but it can also be realized in an experiment, e.g., by applying a magnetic field of the form $\vec{B} = \frac{B_0}{\sqrt{2}}(\hat{x} + \hat{z})$.

**CQS 1.4** Choose all of the following statements that are true regarding the uncertainty principle and uncertainty of measurement in QM.

I. The uncertainty principle refers to the inability of a measuring apparatus to be infinitely precise.
II. The uncertainty in the measurement of an observable can be determined by making a large number of measurements in identically prepared quantum systems in state $|\chi\rangle$, and calculating the standard deviation of those measurements.
III. The uncertainty in the measurement of an observable can never be zero, because the product of the uncertainties of two observables whose operators do not commute must always be $\geq \frac{\hbar}{2}$.

A. I only   **B. II only**   C. III only   D. All of the above   E. None of the above

*Class discussion for CQS 1.4*
- The uncertainty principle describes the observation that two observables whose operators do not commute (and thus whose operators do not have a complete set of simultaneous eigenstates) can never be measured with 100% certainty in the same state, i.e., the system cannot be in an eigenstate of both operators at the same time.
- The uncertainty in the measurement of an observable must be determined by performing a large number of measurements on identically prepared systems, instead of making repeated measurements on the same system. This is because the first measurement will collapse the state of a system into an eigenstate of the operator corresponding to the observable being measured, which in general will not be the initial state (unless the initial state was an eigenstate of the operator corresponding to the observable being measured).

- Emphasize to students that the generalized uncertainty principle is $\sigma_A{}^2 \sigma_B{}^2 \geq \left(\frac{1}{2i}\langle\chi|[\hat{A},\hat{B}]|\chi\rangle\right)^2$, and that $\sigma_x \sigma_{p_x} \geq \frac{\hbar}{2}$ is only the special case for position and momentum in one dimension.

**CQS 2.1** Choose all of the following statements that are correct about uncertainties in the measurement of different components of spin, for an ensemble of identical systems in an eigenstate of $\hat{S}_z$.

I. $\sigma_{S_z}{}^2 = 0$
II. $\sigma_{S_x}{}^2 \neq 0$
III. $\sigma_{S_y}{}^2 = 0$

A. I only    B. III only    **C. I and II only**    D. I and III only    E. None of the above

**CQS 2.2** Choose all of the following statements that are correct about the uncertainty in the measurement of an observable $A$ in state $|\chi\rangle$ (which is not an eigenstate of $\hat{A}$).

I. $\sigma_A^2 = \langle(A - \langle A\rangle)\rangle^2$
II. $\sigma_A^2 = \langle A^2\rangle - \langle A\rangle^2$
III. $\sigma_A^2 = \langle(A - \langle A\rangle)^2\rangle$

A. II only    B. I and II only    C. I and III only    **D. II and III only**    E. All of the above

**CQS 2.3** *Consider the uncertainty $\sigma_{S_x}$ in the observable $S_x$ in a given quantum state $|\chi\rangle$. Choose all the following statements that are correct. [Please give students sufficient time for this question to perform the necessary calculations.]

Note:
$\sigma_{S_x}^2 = \langle\hat{S}_x{}^2\rangle - \langle\hat{S}_x\rangle^2$    $\langle\hat{S}_x\rangle = \langle\chi|\hat{S}_x|\chi\rangle$    $\hat{S}_x = \frac{\hbar}{2}\begin{pmatrix}0 & 1\\ 1 & 0\end{pmatrix}$    $\hat{S}_x^2 = \frac{\hbar^2}{4}\begin{pmatrix}1 & 0\\ 0 & 1\end{pmatrix}$

I. $\sigma_{S_x} = 0$ in the state $|z\rangle$.
II. $\sigma_{S_x} = \frac{3}{5}\left(\frac{\hbar}{2}\right)$ in the state $\frac{1}{\sqrt{5}}|+z\rangle + \frac{2}{\sqrt{5}}|-z\rangle \doteq \frac{1}{\sqrt{5}}\begin{pmatrix}1\\2\end{pmatrix}$.
III. $\sigma_{S_x} = \frac{1}{5}\left(\frac{\hbar}{2}\right)$ in the state $\frac{1}{\sqrt{5}}|+z\rangle + \frac{2}{\sqrt{5}}|-z\rangle \doteq \frac{1}{\sqrt{5}}\begin{pmatrix}1\\2\end{pmatrix}$.

**A. II only**    B. III only    C. I and II only    D. I and III only    E. None of the above

**CQS 2.4** *Consider the state $|\chi\rangle = \frac{1}{\sqrt{5}}|+z\rangle + \frac{2}{\sqrt{5}}|-z\rangle$. You make a measurement of $S_z$ in this state and obtain $\frac{\hbar}{2}$. Choose all of the following statements that are true.

Note:
$\sigma_{S_z}^2 = \langle\hat{S}_z{}^2\rangle - \langle\hat{S}_z\rangle^2$    $\langle\hat{S}_z\rangle = \langle\chi|\hat{S}_z|\chi\rangle$    $\hat{S}_z^2 = \hat{S}_x^2 = \frac{\hbar^2}{4}\begin{pmatrix}1 & 0\\ 0 & 1\end{pmatrix}$
$\hat{S}_z = \frac{\hbar}{2}\begin{pmatrix}1 & 0\\ 0 & -1\end{pmatrix}$    $\hat{S}_x = \frac{\hbar}{2}\begin{pmatrix}0 & 1\\ 1 & 0\end{pmatrix}$

I. A measurement of $S_x$ made in immediate succession will have uncertainty $\sigma_{S_x} = \frac{3}{5}\left(\frac{\hbar}{2}\right)$.
II. A measurement of $S_x$ made in immediate succession will have uncertainty $\sigma_{S_x} = \frac{\hbar}{2}$.
III. A measurement of $S_x$ made in immediate succession will have uncertainty $\sigma_{S_x} = 0$.

A. I only    **B. II only**    C. III only    D. None of the above    E. Not enough information

*Note: CQS 2.3-2.4 were added to the CQS for the in-person implementation to give students practice with numerically calculating measurement uncertainty.*

**CQS 2.5**   The generalized uncertainty principle for two operators is $\sigma_A^2 \sigma_B^2 \geq \left(\frac{1}{2i}\langle[\hat{A},\hat{B}]\rangle\right)^2$. For $\hat{S}_x$ and $\hat{S}_y$, this means $\sigma_{S_x}^2 \sigma_{S_y}^2 \geq \left(\frac{1}{2i}\langle[\hat{S}_x,\hat{S}_y]\rangle\right)^2$. Choose all of the following statements that are correct for a system consisting of a spin-1/2 particle.
  I.   If the state of the system is $|z\rangle$, we have $\langle[\hat{S}_x,\hat{S}_y]\rangle = \langle+z|[\hat{S}_x,\hat{S}_y]|+z\rangle = 0$.
  II.  If the state of the system is $|y\rangle$, we have $\langle[\hat{S}_x,\hat{S}_y]\rangle = \langle+y|[\hat{S}_x,\hat{S}_y]|+y\rangle = 0$.
  III. If we measure the observables $S_x$ or $S_y$ in the state $|+y\rangle$, we will measure $\frac{\hbar}{2}$ in each case with 100% certainty.

**A. II only**    B. I and II only    C. II and III only    D. All of the above
E. None of the above

*Class discussion for CQS 2.1-2.5*
- Emphasize that the uncertainty in the measurement of an observable depends on both the observable being measured and the state of a given system.
- Note that the uncertainty in the measurement of a component of spin can in fact be zero for choice II, $\sigma_{S_x}^2 \sigma_{S_y}^2 = \left(\frac{1}{2i}\langle+y|[\hat{S}_x,\hat{S}_y]|+y\rangle\right) = 0$, because the commutator $[\hat{S}_x,\hat{S}_y] = i\hbar\hat{S}_z$ and the expectation value of $S_z$ in the state $|+y\rangle$ is $\langle+y|\hat{S}_z|+y\rangle = 0$. At the same time, $\sigma_{S_y}^2 = 0$ in an eigenstate of $\hat{S}_y$, and thus $\sigma_{S_x}^2 \sigma_{S_y}^2 = \frac{1}{2i}(\langle+y|[\hat{S}_x,\hat{S}_y]|+y\rangle)^2$ reduces to $0 = 0$, which is true but not useful. There are other ways to show this as well. In summary, to calculate $\langle[\hat{S}_x,\hat{S}_y]\rangle$:
  - If the state of the system is $|+z\rangle$, we have $\langle[\hat{S}_x,\hat{S}_y]\rangle = \langle+z|[\hat{S}_x,\hat{S}_y]|+z\rangle = i\frac{\hbar^2}{2}$.
  - If the state of the system is $|+y\rangle$, we have $\langle[\hat{S}_x,\hat{S}_y]\rangle = \langle+y|[\hat{S}_x,\hat{S}_y]|+y\rangle = 0$.

**CQS 3.1**   Consider the following pairs of operators, which correspond to observables. Which of these pairs of operators can have simultaneous eigenstates?
Note: $[\hat{S}_y, \hat{S}^2] = 0$
  I.   $\hat{S}_y$ and $\hat{S}_x$
  II.  $\hat{S}_y$ and $(\hat{S}_x^2 + \hat{S}_z^2) = (\hat{S}^2 - \hat{S}_y^2)$
  III. $\hat{S}_z$ and $(C\hat{S}_x + \hat{S}_z^2)$, where $C$ is an appropriate constant.

**A. II only**    B. I and II only    C. I and III only    D. II and III only    E. None of the above

**CQS 3.2**   Consider the following pairs of observables. Do quantum states exist for which the measurements of both observables in a system will yield definite values with 100% certainty? Select the pairs for which the answer is yes.
  I.   $S_y$ and $(S_x^2 + 2S_y^2 + S_z^2) = (S^2 + S_y^2)$
  II.  $S_y$ and $(S_x + S_y + S_z)$
  III. $S_y$ and $(CS_y + S_xS_y^2)$, where $C$ is an appropriate constant.

**A. I only**    B. II only    C. III only    D. I and II only    E. None of the above

*Note*: In a spin-1/2 system only, $\hat{S}_x^2 = \hat{S}_y^2 = \hat{S}_z^2 = \frac{\hbar^2}{4}\hat{I}$. This property is not important for answering the questions CQS 3.1-3.2, which are generalizable to all spin systems. The property can, however lead to behavior such as $[\hat{S}_x, \hat{S}_x^2 + \hat{S}_z^2] = 0$, which is not generalizable.

**CQS 3.3** Choose all of the following statements that are correct.
I. $[\hat{S}_x, \hat{S}_y \hat{S}_x] = \hat{S}_x[\hat{S}_x, \hat{S}_y]$
II. $[\hat{S}_x, \hat{S}_y \hat{S}_x] = \hat{S}_y[\hat{S}_x, \hat{S}_x]$
III. $[\hat{S}_x, C_1\hat{S}^2 + C_2\hat{S}_z] = C_2[\hat{S}_x, \hat{S}_z]$ (where $C_1$ and $C_2$ are non-zero constants)
A. I only    B. II only    **C. III only**    D. I and III only    E. II and III only

**CQS 3.4** \*\*The operators $\hat{S}_x$ and $\hat{S}^2$ commute. Suppose you measured $S_x$ in an initial state, in a spin-1/2 system, and obtained a value of $\frac{\hbar}{2}$. What will a subsequent measurement of $S^2$ yield?

A. $\frac{\hbar}{2}$   B. Either $\frac{\hbar^2}{4}$ or $-\frac{\hbar^2}{4}$   C. $\frac{\hbar^2}{4}$   **D. $\frac{3\hbar^2}{4}$**   E. Either $\frac{3\hbar^2}{4}$ or $-\frac{3\hbar^2}{4}$

**CQS 3.5** \*\*The operators $\hat{S}_x$ and $\hat{S}^2$ commute. Suppose you measure $S_x$ in some initial state, for a spin-1/2 system, and obtain a value of $\frac{\hbar}{2}$. What will the state be if $S^2$ is measured in immediate succession?
**A. $|+x\rangle$**
B. Either $|+x\rangle$ or $|-x\rangle$
C. The system can be in any of the eigenstates of $\hat{S}_x$, $\hat{S}_y$, or $\hat{S}_z$
D. The system can be in any state in the Hilbert space after the measurement, since every state is an eigenstate of $\hat{S}^2$.
E. Not enough information

*\*\*Note: CQS 3.4-3.5 were added to the CQS for the in-person implementation after observing the student difficulties present after the online implementation.*

*Class discussion for CQS 3.1-3.5*
- *Discuss how the operator $\hat{S}^2 = \hat{S}_x^2 + \hat{S}_y^2 + \hat{S}_z^2$ commutes with all three operators $\hat{S}_x$, $\hat{S}_y$, and $\hat{S}_z$.*
  - *Note that neither $\hat{S}_x$ nor $\hat{S}_y$ would commute with an operator $\hat{S}_x^2 + \hat{S}_y^2$. However, $\hat{S}_x$ would commute with $\hat{S}_y^2 + \hat{S}_z^2$, which can be alternatively expressed as $\hat{S}^2 - \hat{S}_x^2$. An analogous case holds for $\hat{S}_y$.*
  - *The situation is special for $\hat{S}^2$ for spin-1/2 systems, because $\hat{S}_x^2$, $\hat{S}_y^2$, and $\hat{S}_z^2$ are all proportional to the identity operator $\hat{I}$, such that, e.g., $\hat{S}_x$ does commute with $\hat{S}_x^2 + \hat{S}_y^2$.*
- *Discuss the connection between a complete set of simultaneous eigenstates of commuting operators and the implications for the measurements of commuting observables in immediate succession. (Students may need help seeing the connection between CQS 3.1-3.2.)*
  - *Emphasize to students the implications of degeneracy in the eigenvalue spectrum of $\hat{S}^2$, i.e., every eigenstate of $\hat{S}_x$ is an eigenstate of $\hat{S}^2$, but not the other way around.*
- *Emphasize also that $\hat{S}_x$, $\hat{S}_y$, and $\hat{S}_z$ are on the same footing with regard to commutation relations to one another and to $\hat{S}^2$.*
  - *It is conventional to write $\hat{S}_x, \hat{S}_y, \hat{S}_z$ for a spin-1/2 system (which are proportional to the Pauli matrices) in a basis consisting of eigenstates of $\hat{S}_z$ (but this is like defining the x-axis in classical mechanics to be horizontal).*

- Any one of these, for example, $\hat{S}_z$, is "special" for a system only if the Hamiltonian of the system commutes with $\hat{S}_z$.

**CQS 4.1** ***Consider the Hermitian operators $\hat{A}$ and $\hat{B}$, which correspond to observables $A$ and $B$, respectively. They are *compatible* operators ($[\hat{A}, \hat{B}] = 0$). The Hamiltonian is given by $\hat{H} = \hat{A} + \hat{B}$. All three operators have non-degenerate eigenstates. Suppose you measure the observable $A$ and obtain $A_0$. After this measurement, the state is definitely an eigenstate of which of the following operators?

I. $\hat{A}$
II. $\hat{B}$
III. $\hat{H}$

A. I only   B. I and II only   C. I and III only   **D. All of the above**   E. None of the above

**CQS 4.2** ***Consider the Hermitian operators $\hat{C}$ and $\hat{D}$, which correspond to observables. They are *incompatible* operators ($[\hat{C}, \hat{D}] \neq 0$). The Hamiltonian is given by $\hat{H} = \hat{C} + \hat{D}$. Suppose you measure $C$ and obtain $C_0$. After this measurement, the state is definitely an eigenstate of which of the following operators?

I. $\hat{C}$
II. $\hat{D}$
III. $\hat{H}$

**A. I only**   B. I and II only   C. I and III only   D. All of the above   E. None of the above

*\*\*\*Note: For the online implementation, the choices for CQS 4.1-4.2 read as follows. (The operators in the preceding question CQS 4.2 were renamed $\hat{C}$ and $\hat{D}$ for the in-person implementation to avoid confusion.)*
Consider the Hermitian operators $\hat{A}$ and $\hat{B}$, which correspond to observables. They are compatible [CQS 4.1] / incompatible [CQS 4.2] operators. The Hamiltonian is given by $\hat{H} = \hat{A} + \hat{B}$. Suppose you measure energy and obtain $E_0$. Choose all of the following that are correct after the measurement of energy:
I. A is well-defined.
II. B is well-defined.
III. The state collapses to an eigenstate of $\hat{B}$ immediately after the measurement of energy.
CQS 4.1:   A. I only   B. III only   C. I and II only   D. II and III only   **E. All of the above**
CQS 4.2:   A. I only   B. III only   C. I and II only   D. II and III only   **E. None of the above**

**CQS 4.3** Choose all of the following statements that are correct if $\hat{A}$ and $\hat{B}$ are *incompatible* operators with non-degenerate eigenstates.

I. It is impossible to find a complete set of simultaneous eigenstates for $\hat{A}$ and $\hat{B}$.
II. In a given quantum state, for three successive measurements A→B→A (*assuming no time evolution of the state has taken place*), the two measurements of A must yield the same value.
III. It is possible to infer the value of the observable $B$ after the measurement of the observable $A$ returns a particular value for $A$.

**A. I only**   B. III only   C. I and II only   D. All of the above   E. None of the above

[1] P. Lebedev, C. Lindstrøm, and M. D. Sharma, Making linear multimedia interactive: questions, solutions and types of reflection, Eur. J. Phys. **42**, 015707 (2020).
[2] M. C. Wittmann, J. T. Morgan, and L. Bao, Addressing student models of energy loss in quantum tunnelling, Eur. J. Phys. **26**, 939 (2005).
[3] A. Johansson, Undergraduate quantum mechanics: lost opportunities for engaging motivated students?, Eur. J. Phys. **39**, 025705 (2018).
[4] L. G. de Peralta, L. A. Poveda, and B. Poirier, Making relativistic quantum mechanics simple, Eur. J. Phys. **42**, 055404 (2021).
[5] E. B. Rodríguez, L. M. A. Aguilar, and E. P. Martínez, A full quantum analysis of the Stern–Gerlach experiment using the evolution operator method: analyzing current issues in teaching quantum mechanics, Eur. J. Phys. **38**, 025403 (2017).
[6] P. Hadzidaki, G. Kalkanis, and D. Stavrou, Quantum mechanics: a systemic component of the modern physics paradigm, Phys. Educ. **35**, 386 (2000).
[7] M. Chhabra and R. Das, Quantum mechanical wavefunction: visualization at undergraduate level, Eur. J. Phys. **38**, 015404 (2016).
[8] I. D. Johnston, K. Crawford, and P. R. Fletcher, Student difficulties in learning quantum mechanics, Int. J. Sci. Educ. **20**, 427 (1998).
[9] R. Müller and H. Wiesner, Teaching quantum mechanics on an introductory level, Am. J. Phys. **70**, 200 (2002).
[10] M. C. Wittmann, R. N. Steinberg, and E. F. Redish, Investigating student understanding of quantum physics: Spontaneous models of conductivity, Am. J. Phys. **70**, 218 (2002).
[11] C. Singh, M. Belloni, and W. Christian, Improving students' understanding of quantum mechanics, Phys. Today **59**, 43 (2006).
[12] C. Singh, Student understanding of quantum mechanics at the beginning of graduate instruction, Am. J. Phys. **76**, 277 (2008).
[13] G. Zhu and C. Singh, Surveying students' understanding of quantum mechanics in one spatial dimension, Am. J. Phys. **80**, 252 (2012).
[14] E. Marshman and C. Singh, Framework for understanding the patterns of student difficulties in quantum mechanics, Phys. Rev. ST Phys. Educ. Res. **11**, 020119 (2015).
[15] C. Singh and E. Marshman, Review of student difficulties in upper-level quantum mechanics, Phys. Rev. ST Phys. Educ. Res. **11**, 020117 (2015).
[16] V. Dini and D. Hammer, Case study of a successful learner's epistemological framings of quantum mechanics, Phys. Rev. Phys. Educ. Res. **13**, 010124 (2017).
[17] C. Singh, Student understanding of quantum mechanics, Am. J. Phys. **69**, 885 (2001).
[18] C. Singh, Student difficulties with quantum mechanics formalism, AIP Conf. Proc. **883**, 185 (2007).
[19] E. Gire and E. Price, Structural features of algebraic quantum notations, Phys. Rev. ST Phys. Educ. Res. **11**, 020109 (2015).
[20] D. F. Styer, Common misconceptions regarding quantum mechanics, Am. J. Phys. **64**, 31 (1996).
[21] C. Singh and G. Zhu, Cognitive issues in learning advanced physics: An example from quantum mechanics, AIP Conf. Proc. **1179**, 63 (2009).
[22] D. Domert, C. Linder, and Å. Ingerman, Probability as a conceptual hurdle to understanding one-dimensional quantum scattering and tunnelling, Eur. J. Phys. **26**, 47 (2004).

[23]   G. Zhu and C. Singh, Improving students' understanding of quantum measurement. I. Investigation of difficulties, Phys. Rev. ST Phys. Educ. Res. **8**, 010117 (2012).
[24]   G. Passante, P. J. Emigh, and P. S. Shaffer, Examining student ideas about energy measurements on quantum states across undergraduate and graduate levels, Phys. Rev. ST Phys. Educ. Res. **11**, 020111 (2015).
[25]   C. Singh, Transfer of learning in quantum mechanics, AIP Conf. Proc. **790**, 23 (2005).
[26]   T. Kaur, D. Blair, J. Moschilla, and M. Zadnik, Teaching Einsteinian physics at schools: part 2, models and analogies for quantum physics, Phys. Educ. **52**, 065013 (2017).
[27]   P. Jolly, D. Zollman, N. S. Rebello, and A. Dimitrova, Visualizing motion in potential wells, Am. J. Phys. **66**, 57 (1998).
[28]   D. A. Zollman, N. S. Rebello, and K. Hogg, Quantum mechanics for everyone: Hands-on activities integrated with technology, Am. J. Phys. **70**, 252 (2002).
[29]   A. Kohnle, M. Douglass, T. J. Edwards, A. D. Gillies, C. A. Hooley, and B. D. Sinclair, Developing and evaluating animations for teaching quantum mechanics concepts, Eur. J. Phys. **31**, 1441 (2010).
[30]   A. Kohnle, I. Bozhinova, D. Browne, M. Everitt, A. Fomins, P. Kok, G. Kulaitis, M. Prokopas, D. Raine, and E. Swinbank, A new introductory quantum mechanics curriculum, Eur. J. Phys. **35**, 015001 (2013).
[31]   E. Marshman and C. Singh, Validation and administration of a conceptual survey on the formalism and postulates of quantum mechanics, Phys. Rev. Phys. Educ. Res. **15**, 020128 (2019).
[32]   G. Zhu and C. Singh, Improving students' understanding of quantum measurement. II. Development of research-based learning tools, Phys. Rev. ST Phys. Educ. Res. **8**, 010118 (2012).
[33]   E. Marshman and C. Singh, Investigating and improving student understanding of the probability distributions for measuring physical observables in quantum mechanics, Eur. J. Phys. **38**, 025705 (2017).
[34]   E. Marshman and C. Singh, Investigating and improving student understanding of the expectation values of observables in quantum mechanics, Eur. J. Phys. **38**, 045701 (2017).
[35]   C. Singh, Interactive learning tutorials on quantum mechanics, Am. J. Phys. **76**, 400 (2008).
[36]   E. Marshman and C. Singh, Interactive tutorial to improve student understanding of single photon experiments involving a Mach–Zehnder interferometer, Eur. J. Phys. **37**, 024001 (2016).
[37]   E. Mazur, *Peer Instruction: A User's Manual* (Prentice Hall, Upper Saddle River, N.J., 1997).
[38]   C. H. Crouch and E. Mazur, Peer Instruction: Ten years of experience and results, Am. J. Phys. **69**, 970 (2001).
[39]   P. Justice, E. Marshman, and C. Singh, Improving student understanding of quantum mechanics underlying the Stern–Gerlach experiment using a research-validated multiple-choice question sequence, Eur. J. Phys. **40**, 055702 (2019).
[40]   P. Justice, E. Marshman, and C. Singh, Student understanding of Fermi energy, the Fermi–Dirac distribution and total electronic energy of a free electron gas, Eur. J. Phys. **41**, 015704 (2019).
[41]   C. Singh and G. Zhu, Improving students' understanding of quantum mechanics by using peer instruction tools, AIP Conf. Proc. **1413**, 77 (2012).

[42]   P. Hu, Y. Li, and C. Singh, Challenges in addressing student difficulties with time-development of two-state quantum systems using a multiple-choice question sequence in virtual and in-person classes, Eur. J. Phys. **43**, 025704 (2022).
[43]   P. Justice, E. Marshman, and C. Singh, Development, validation and in-class evaluation of a sequence of clicker questions on Larmor precession of spin in quantum mechanics, *Physics Education Research Conference Proc.* (2019), pp. 269–274.
[44]   P. Justice, E. Marshman, and C. Singh, Development and validation of a sequence of clicker questions for helping students learn addition of angular momentum in quantum mechanics, Phys. Rev. ST Phys. Educ. Res. **9**, 010101 (2013).
[45]   Y. Li, A. Kohnle, and G. Passante, Student difficulties with quantum uncertainty in the context of discrete probability distributions, *Physics Education Research Conference Proc.* (2021), pp. 227–232.
[46]   J. H. McMillan and J. F. Wergin, *Understanding and Evaluating Educational Research* (Prentice-Hall, Inc, 1998).
[47]   R. R. Hake, Interactive-engagement versus traditional methods: A six-thousand-student survey of mechanics test data for introductory physics courses, Am. J. Phys. **66**, 64 (1998).
[48]   J. Cohen, *Statistical Power Analysis for the Behavioral Sciences* (Academic Press, 2013).